\documentclass[journal]{IEEEtran}

\usepackage{cite}
\usepackage{url}

\usepackage{graphics} 
\usepackage{epsfig} 
\usepackage{amsmath} 
\usepackage{amssymb}  

\usepackage{amsfonts}

\usepackage{multirow}
\usepackage[linesnumbered,ruled,vlined,Algorithm ]{algorithm }
\usepackage{algpseudocode}

\usepackage{hyperref}
\usepackage{epstopdf}

\usepackage{array}
\usepackage{enumerate}
\usepackage{caption}
\usepackage{subcaption}
\usepackage{stfloats}

\usepackage[english]{babel} 
\usepackage{amsfonts,amsthm,bm} 
 
\usepackage{cleveref}

 \usepackage{cite}

\begin{document}
\thispagestyle{empty}
\onecolumn 
\vspace*{\fill}
This work submitted to journal/IEEE transaction for possible publication. Copyright may be transferred without notice, after which this version may no longer be accessible.
\vspace*{\fill}

\twocolumn
\newpage 
%
\title{ A Decentralized Multi-UAV Spatio-Temporal Multi-Task Allocation Approach for Perimeter Defense  }

%
%
%

\author{Shridhar~Velhal,~Suresh~Sundaram,  
and~Narasimhan~Sundararajan 

\thanks{ Shridhar Velhal is a PhD student at Department of Aerospace Engineering, Indian Institute of Science, Bengaluru, India.
       {\tt\small velhalb@iisc.ac.in}}%
\thanks{ Suresh Sundaram is an Associate Professor at Department of Aerospace Engineering, Indian Institute of Science, Bengaluru, India.
        {\tt\small vssuresh@iisc.ac.in}}%

\thanks{  Narasimhan Sundararajan  is with the School of Electrical and
Electronics Engineering, Nanyang Technological University, Singapore {\tt\small ensundara@ntu.edu.sg}}%
 }

%
%

\markboth{}
{  }

%



\maketitle
\begin{abstract}
This paper provides a new solution approach to a multi-player perimeter defense game, in which the intruders' team tries to enter the territory, and a team of defenders protects the territory by capturing intruders on the perimeter of the territory. The objective of the defenders is to detect and capture the intruders before the intruders enter the territory. Each defender independently senses the intruder and computes his trajectory to capture the assigned intruders in a cooperative fashion. The intruder is estimated to reach a specific location on the perimeter at a specific time. Each intruder is viewed as a spatio-temporal task, and the defenders are assigned to execute these spatio-temporal tasks. At any given time, the perimeter defense problem is converted into a Decentralized Multi-UAV Spatio-Temporal Multi-Task Allocation (DMUST-MTA) problem. The cost of executing a task for a trajectory is defined by a composite cost function of both the spatial and temporal components. In this paper, a decentralized consensus-based bundle algorithm has been modified to solve the spatio-temporal multi-task allocation problem and the performance evaluation of the proposed approach is carried out based on Monte-Carlo simulations. The simulation results show the effectiveness of the proposed approach to solve the perimeter defense game under different scenarios. Performance comparison with a state-of-the art centralized approach with full observability, clearly indicate that DMUST-MTA achieves  similar performance in a decentralized way with partial observability conditions with a lesser computational time and easy scaling up.  
\end{abstract}


%

\section{Introduction}

\IEEEPARstart{R}{apidly} evolving technologies in the autonomous operation of Unmanned Aerial Vehicles (UAVs) and associated developments in low-cost sensors have created significant interest among researchers in using them for various civil and military applications. In particular, the autonomous aerial vehicles are often used for logistics \cite{kuru2019analysis}, medical emergencies\cite{rosser2018surgical}, agricultural operations \cite{mogili2018review}, security and surveillance applications \cite{harikumar2019mission}, \cite{harikumar2018multi}.  Recent increases in the use of UAVs in an open environment at lower altitudes introduce many challenging problems in safety, privacy, and security \cite{solodov2018analyzing}. 
 
Serious security issues arise when a freely operating UAV intrudes and flies over critical areas such as airports, nuclear facilities, chemical industries, ports, government infrastructure, etc.  Advanced imaging sensors in the UAVs create severe threats to the security of protected areas. Protecting airspace from such a UAV's physical attack is challenging and has significant importance in both civilian and military applications. This challenge mainly consists of two aspects, viz., early detection of the attacker as the first challenge and the second one being neutralizing the attacker. Recently, researchers have been working on using autonomous multi-UAV systems to protect a territory from intruders.  In \cite{ganti2016implementation}, a method has been proposed to track the UAV, and based on its trajectory, one can identify a potential intruder. To neutralize the intruder; one can use the manipulation of its sensors by conducting a cyber-attack on them \cite{kerns2014unmanned}. A survey of counter UAV systems to capture the intruders is presented in \cite{kang2020protect}.

\begin{figure}[!t]
     \centering
    \includegraphics[width=1\linewidth]{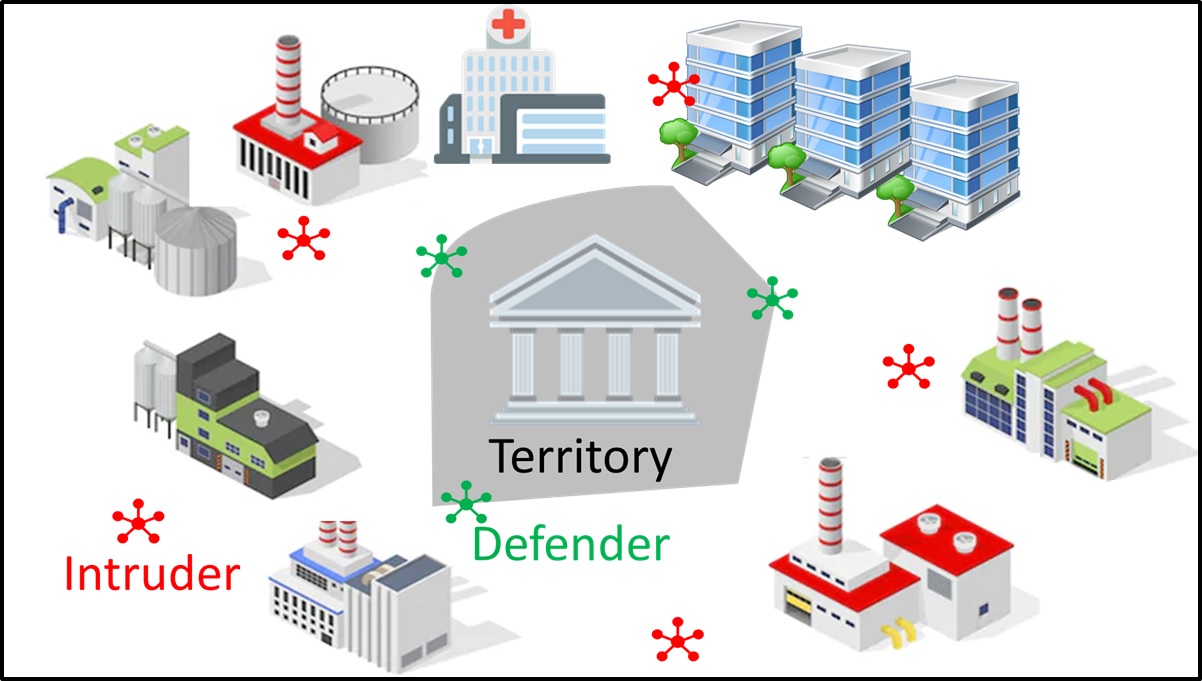}
    \caption{A typical scenario for the protection of critical infrastructure}
    \label{fig:TP1}
\end{figure} 

Figure \ref{fig:TP1} shows a typical scenario for protecting critical infrastructure (nuclear facilities, airports, chemical industries, ports, etc.) from a group of intruders.  The airspace around any critical infrastructure is called a territory, represented using a convex gray region in Fig. 1. There is a set of UAVs operating in the perimeter of the territory (called defenders, illustrated in green colour) and they capture the intruders (represented in red colour) before they enter the region. When an intruder enters the territory, the critical infrastructures' privacy, safety, and security are compromised. The intruder must be captured/neutralized on the perimeter of the territory. Defenders operate only inside the territory. One should note that the number of intruders is not a constant, and it can vary over time. The Perimeter Defense Problem (PDP) is defined as a group of defenders who cooperatively capture the intruders on the perimeter of the territory with the aim of protecting the infrastructure. 
 
A multi-robot perimeter patrol problem for adversarial conditions was proposed in \cite{agmon2008multi}. The non-deterministic patrol algorithm was employed for patrolling the territory in an adversarial setting. The path defense \cite{chen2014path} approach provides a strategy for a 1-to-1 case, where one defender protects the target from one intruder while operating only on a predefined path using a fast-marching method.  This constraint limits the defender's movement, and they have to respond according to the intruders' actions. In a perimeter defense problem \cite{shishika2018local,shishika2020cooperative,shishika2019team, paulos2019decentralization}, intruders are trying to enter the territory, and the defenders will protect the territory while operating only on the perimeter of the territory.  The geometric method \cite{shishika2018local} provides a reachability analysis for the feasible pairing of defenders to the intruders. A maximum matching algorithm has been used for the task allocation among the feasible pairs. The maximum matching algorithm assumes that each defender plays a game independently; this method may lead to a non-capture of the intruder, which was feasible via a 2-vs-1 game.  

In \cite{shishika2020cooperative }, a cooperative defense strategy has been proposed to handle PDP. First, a local game region is defined that forms a team of defenders who can capture multiple intruders in that local region. Here, the defenders need to cooperate with other agents in the local region to neutralize the intruder. 

A perimeter defense problem with sensing by the patroller UAVs are addressed in \cite{shishika2019team}; the paper provides analytical conditions on the number of patrollers and defenders required to defend a territory from  intruders. It also provides a bound on inter-agent separation for better detection of an intruder. The system with decentralized sensing is handled using a policy decentralization using reinforcement learning \cite{paulos2019decentralization}. The policy is leaned using a centralized expert solution with full communication and a condition on defender moving either clockwise or anti-clockwise.  The action space for the defender is limited to only two actions; move clockwise or anti-clockwise. The reinforcement learning-based solution gets complicated when the defender moves in any direction.  
 
A complete review of research in the area of perimeter defense games is presented in \cite{shishika2020review}. The review paper also points out the limitations of the current works on the defender dynamics, sequential capture, fast intruders, and partial information. Further, the above-mentioned works employ only a centralized solution for PDP and are not scalable. Hence there is a need to develop a decentralized approach to overcome the above-mentioned limitations and for easy scaling up and this paper attempts to address this issue.
 
In this paper, we present a consensus-based decentralized multi-UAV spatio-temporal multi-task allocation (DMUST-MTA) approach to solve the perimeter defense problem with sequential capture of varying number of intruders under partial observability conditions. The proposed method can handle a larger number of intruders using a smaller number of defenders.  In a typical perimeter defense problem, the defender UAVs operate inside the territory and capture the intruders on the perimeter. Defenders are equipped with sensors to detect the intruders and cooperate with other defenders to capture the intruders. The problem of capturing a varying   number of intruders is first converted into a spatio-temporal multi-task allocation problem. The estimated arrival times and the locations of the intruders at any given time are used to assign the spatio-temporal tasks. These spatio-temporal multiple tasks are allocated to the defenders using a decentralized spatio-temporal multi-task allocation approach. The one-dimensional multiple-task allocation using a Consensus-Based Bundle Algorithm (CBBA) presented in  \cite{choi2009consensus} is extended here to handle the spatio-temporal multiple-task allocation problem. 
 
The modified CBBA uses a decentralized auction method where every defender bid for neutralizing the observed intruders. The bidding cost is computed for every defender's chosen trajectory, and a composite loss function is then used to unite both the spatial and temporal components. A consensus is then formed based on partial communications among the defenders. 

The output of this task allocation scheme defines a trajectory for each defender to neutralize the intruders. The performance of the proposed DMUST-MTA has been evaluated using numerical simulations. Monte-Carlo simulations have also been conducted by varying the speed of defenders, sensing radius, and temporal separation between the intruders. As the speed of defenders increases, the success percentage of neutralizing the intruders  increases. The increasing sensing radius provides more planning time for the defender, and hence this also increases success percentage. Similarly, the temporal separations between the intruders increase the choices to capture the intruders, hence, it increases the chances of success. The performance of DMUST-MTA is compared with a centralized local game region (LGR) cooperative approach \cite{shishika2020cooperative} under full observability condition and a one-to-one setting. The results clearly indicate that both the methods achieved similar performances, but DMUST-MTA requires lesser computational time. The computational time increases exponentially for LGR when the number of defenders increases, whereas DMUST-MTA is decentralized with a lesser computationally intensive approach, and it can handle partial observability.  The main contributions of this paper are as follows.

 \begin{enumerate} 
    \item  Formalization of the perimeter defense problem into a multi-UAV spatio-temporal multi-task allocation problem. The presence of both the spatial and temporal dimensions and dynamic environments makes the solution for MUST-MTA challenging. 
    \item  A composite loss function is defined to handle the spatio-temporal nature of the task. The formulated spatio-temporal multi-task allocation problem is solved using modified consensus-based distributed multi-task allocation algorithm. 
    \item  The perimeter defence problem has been extended for limiting cases where the number of defenders is lower than the number of intruders.
\end{enumerate}
  
The paper is organized as follows; Section \ref{sec:related works} provides a review of the territory protection problem and task allocation methods. Section \ref{sec:STMTA for PDP} presents the mathematical formulation of the perimeter defense problem and shows how one can convert this PDP problem into a spatio-temporal multi-task allocation problem. Section \ref{sec:DMUST-MTA} presents a decentralized multi-UAV spatio-temporal  multi-task allocation algorithm to capture a varying number of intruders.  Monte-Carlo simulation study results   are presented for the performance evaluation of the proposed approach in Section  \ref{sec:results}. Finally, Section \ref{sec:conclusion} presents the conclusions.

\section{Related Works  } \label{sec:related works} 
In this section, the   related works in the area of Perimeter Defence Problem   are summarised under two categories namely, a) territory protection games and b) Task allocation approach.

\subsection{ Territory Protection Games }  \label{subsec:TPgame} 

Multi-player pursuit-evasion games are crucial for addressing decision-making problems in developing cooperative control of multi-agent systems. The fundamentals of pursuit-evasion games and their challenges were given in the seminal work \cite{isaacs1999differential}. Multiple pursuers and one evader game \cite {bakolas2010optimal, von2018pursuit, von2019multi} provides a cooperative solution for the team of pursuers to capture an evader using a geometric approach based on Voronoi diagrams. A distributed algorithm for multiple pursuers to capture multiple evaders has been demonstrated in \cite{pierson2016intercepting}.   A distributed algorithm for capturing high-speed evader by multiple pursuers has been provided in \cite{lee2020perimeter}. An inverse scenario where a pursuer captures multiple evaders within a circle is discussed in \cite{bopardikar2009cooperative}. This work addresses a challenging open problem where number of evaders is more than the number of pursuers, but evaders were assumed to be uniformly distributed and move radially outward at a constant speed.  

A border defense problem \cite{garcia2020multiple,salmon2020single}, where a pursuer captures an evader before the evader leaves the territory, is solved using game theory based on Apollonius circle.  Security games \cite{tambe2011security, yang2013improving,bucarey2019coordinating} provide resource allocation-based solutions to defenders protecting the paths of agents. In the security games, a defender should be at-least in the vicinity of a static/dynamic agent (also called a target) for its protection.  The agent's path is assumed to be known or scheduled in coordination with the defender; the target and the defender cooperatively protect the target from the attackers. 

The multiple pursuers multiple evaders\ game has been addressed using social spider optimization (SSO) \cite{husodo2020enhanced}. This approach assumes that intruders are static, and the solution is intractable.  

When a pursuer tries to protect a territory from an evader, the pursuit-evasion game becomes a guarding game. A differential game between an attacker and a guard \cite{liang2019differential, garcia2015cooperative, garcia2019cooperative} provides an analytical solution based on the initial position of the attacker. A geometrical capture region is computed analytically based on the location of guards and the territory. If the attacker lies in the capture region, the guard can capture the attacker outside the territory irrespective of its strategy. The guarding game has been studied using different methods such as  fuzzy logic \cite{hsia1993first, lee2002strategy}, and reinforcement learning \cite {analikwu2016reinforcement, raslan2016learning, analikwu2017multi}.  

In the guarding game, if an intruder reaches the target, then the defender team loses the game. However, in the multiplayer game, this is an essential condition for the game, and more often, the defender loses the game; the defender should try to minimize the number of intrusions and capture a maximum number of intruders outside the target region. A reach-avoid game is a game in which intruders(evaders) try to reach the target. The defender(pursuers) tries to avoid this by capturing the intruder before the intruder reaches the target. The intruders get points when an intruder reaches the target, and the defenders get points when it captures the intruder. In the reach-avoid game, both teams try to maximize their effectiveness. A multiplayer reach-avoid game is a numerically intractable problem; \cite{chen2017multiplayer} provides upper bounds for the number of attackers who reaches the target.  The reach-avoid problems with time-varying dynamics, targets, and constraints have been addressed in \cite{fisac2015reach} by extending Hamilton-Jacobi methods. Two defender and one intruder games have been solved using barrier functions based on the shape of the territory, and the defenders' initial positions \cite{yan2019reach, yan2020task}. For a multiplayer game, defenders are allocated to the intruders based on reachable sets obtained from the barrier functions. The problem is solved in two parts; the first is to find a feasible reachable set, and the second is to solve an assignment problem that maximizes the score. In \cite{yan2020guarding} the subspace guarding problem with two fast defenders and one attacker is solved analytically using barrier functions.   
 
A guarding problem for dynamic territory is addressed in an escort team problem. The escort team tries to navigate the payload by positioning escorts around the payload safely.  A deep reinforcement learning-based defensive escort team is proposed to avoid active collision \cite{garg2019defensive}.  As obstacles are repelled by escorting agents, agents will need to position them so that they repel the obstacles on route cooperatively. A learning-based solution is specific to the shape of the payload (trajectory) and, sensing and repelling range of the escorts.

\subsection{Task Allocation Methods}

One of the crucial challenges in using multi-UAV systems for real-world applications is to solve a complex task allocation problem in an unknown/uncertain environment. The objective here is to find an optimal strategy that will assign a set of tasks to the UAVs such that the multi-UAV system achieves its mission successfully. The taxonomy of the task allocation problem and a detailed review of various task allocation schemes can be found in \cite{korsah2013comprehensive,khamis2015multi}. Most of these works address the dynamic allocation of spatially located tasks using any one of the following strategies, viz.,  market-driven strategies \cite{jones2007learning}, game-theoretic strategies \cite{Cui2013}, Hungarian method \cite{chopra2014heterogeneous,Chopra2017} and consensus-based task allocation methods  \cite{choi2009consensus,Brunet2008,zlot2006market,fanti2018decentralized}. 

Recently, in\cite{amador2014dynamic,nunes2017taxonomy,nelke2020market}, the task allocation problems where the spatial task requires a certain time to complete the task (time-window constraints) were presented. The spatial task with a time window constraint is solved using heuristic methods, where a penalty is imposed on the delayed execution of tasks. The works mentioned above are not suitable for a Perimeter Defense Problem (PDP) because PDP tasks are dynamic and are available only at specific time instants. The penalty on delayed task execution is not suitable for PDP as a delayed task execution leads to the critical area's defense failure. To overcome the limitations mentioned above, in this paper, we present a new decentralized approach with distributed sensing.

\section{Spatio-Temporal Multi-task Allocation for Perimeter Defense Problem  } \label{sec:STMTA for PDP}

In this section, first we define the perimeter defense problem with distributed sensing and then convert the perimeter defense problem to a spatio-temporal multi-task allocation problem.

\subsection{Problem Definition} 
 
\begin{figure}[hbt!]
    \centering
    \includegraphics[width=0.90\linewidth]{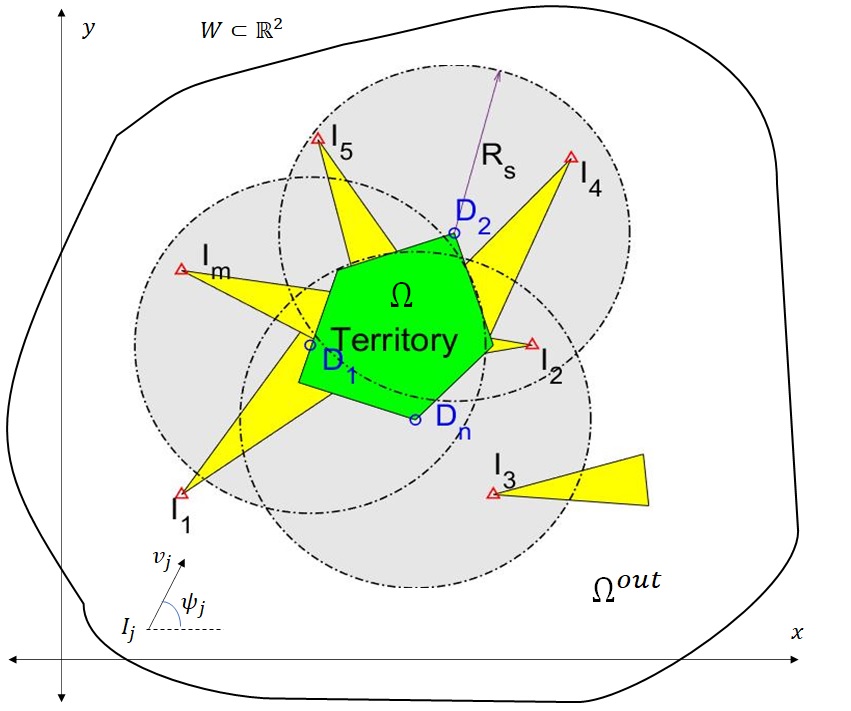}  
    \caption{Perimeter defense problem with the distributed sensing system}
    \label{fig:SenseTask}
\end{figure}
 
The perimeter defense problem aims to design a team of defenders which protects the territory from intruders by capturing them before the intruder enters the territory. The objective of the defender is to detect and capture the intruders before intruders enter the territory. If an intruder enters the territory, then the defense fails. Next, we define the above problem in a mathematical way. 
 
Fig \ref{fig:SenseTask}  shows a typical scenario for a perimeter defense problem with distributed sensing, where each defender can sense the intruder within its sensing range $(R_s)$. The figure highlights the individual defender sensor coverage and also highlights the potential impact region of the intruders. Consider a region $ W \subset R^2$,   a region of interest that includes the critical infrastructure and its neighbourhood. The territory is approximated using a convex shape and is shown by green colour in fig \ref{fig:SenseTask} and is denoted as $ \Omega,  ( \Omega \subset W \subset R^2)$. Let $\partial \Omega $ denotes the territory's perimeter and, ${\Omega}^{out} (W-\Omega)$ denotes the region outside the territory. The length of the perimeter is $L$, and any point on the perimeter is denoted by $s \in [0,L) $. 

Let us consider a set of defenders $ \mathcal{D} = \{ D_1,D_2,\cdots,D_i,\cdots,D_N \} $ operating inside the territory. Defenders are initialized inside the territory, i.e., $D^s (t_0) \in \Omega $. Defenders can communicate synchronously among themselves, but the bandwidth is limited. Each defender is equipped with a homogeneous sensor to detect the intruder's position and velocity within the sensing radius.   Let $D_i^s=(x_D^i  ,y_D^i  )$  be the position of $i^{th}$ defender and velocity $V_D^i$ and heading angle $\psi_D^i$. The kinematic equations of the $i^{th}$ defender are given below,

\begin{align}
\dot{x}_{D}^{i}=V_{D}^{i} \cos \left(\psi_{D}^{i}\right) \quad \dot{y}_{D}^{i}=V_{D}^{i} \sin \left(\psi_{D}^{i}\right), \quad i=1,2, \ldots, N \\
\psi_{D}^{i} \in[0,2 \pi), \quad V_{D}^{i} \in\left[0, V_{D}^{\max }\right], \quad\left(x_{D}^{i}\left(t_{0}\right), y_{D}^{i}\left(t_{0}\right)\right) \in \Omega 
\end{align}

Intruders are initialized outside the territory $I^s (t_0)\in  {\Omega}^{out}$. The number of intruders at any given t is dynamic. Let $M_t$ be the number of intruders that appeared in the ${\Omega}^{out} $ at time instant $t$. Let us denote the set of intruders as   $ \mathcal{I}= \{ I_1,I_2,..., I_j,...,I_{M_t} \}$  Here we assume that where is no coordination between intruders and will not penetrate the territory simultaneously. Let $I_j^s=(x_I^j  ,y_I^j  )$  be the position of $j^{th}$ intruder and velocity $V_I^j$ and heading angle $\psi_I^j$. The intruder's kinematic equations are given below.   
\begin{align}
\dot{x}_{I}^{i}=V_{I}^{i} \cos \left(\psi_{I}^{i}\right) \ \  \dot{y}_{I}^{i}=V_{I}^{i} \sin \left(\psi_{I}^{i}\right), \ \  i=1,2, \cdots,N \\
\psi_{I}^{i} \in[0,2 \pi), \ \ V_{I}^{i} \in\left[0, V_{I}^{\max }\right], \ \  \left(x_{I}^{i}\left(t_{0}\right), y_{I}^{i}\left(t_{0}\right)\right) \in \Omega 
\end{align}

It is assumed that the intruders are nonagile and have a constant heading rate  $ \psi_I^i \le \pm \delta     \  rad/sec $ .
Note, the number of intruders is more than the number of defenders $(M_t  >> N)$. Also, the defender's maximum velocity is more than the intruders' maximum velocity, $V_D^{max}  < V_I^{max}$.  

When the distance between the defender and intruder is less than $ \epsilon $ then, the defender captures the intruder using a safety net mechanism. The capture condition is given as

\begin{equation}
\begin{aligned}
C ={} & \Bigg\{ x \mid \exists i \  \text{ s.t. } \sqrt{\left(x_{I}^{j}-x_{D}^{i}\right)^{2}+\left(y_{I}^{j}-y_{D}^{i}\right)^{2}} \leq \epsilon       \\
 &   \quad  \qquad  \qquad    \& \ 
\left( x_{I}^{j}, y_{I}^{j} \right) \in \Omega^{  out } \Bigg\}
\end{aligned}
\end{equation} 

In \cite{shishika2018local,shishika2020cooperative, shishika2019team, paulos2019decentralization} assumes that the defender neutralizes the intruder using the head-on collision. The capture process will destroy both the defender and intruders and are not practical. In this paper, the defender employs a safety net to capture the intruder, and hence a small number of defenders can protect the territory from a large number of intruders.

From the figure \ref{fig:SenseTask}  , we can observe that $D_1$ can detect only  $I_m$ whereas  $D_2$ can detect $ I_2,I_4$  and $I_5$. Since there exists partial observability in detecting the target, the N defenders cooperatively decide in a decentralized way to capture the intruders. Defenders use low-bandwidth synchronous communication to establish a decentralized decision-making process in neutralizing the intruders. The perimeter defense problem's main objective is to detect the dynamic number of intruders using a distributed sensing approach and allocate the multiple-tasks to different defenders in a decentralized fashion.

\subsection{Spatio-temporal Multi-task Allocation Problem }

\begin{figure*}[!hbt]
\centering
\begin{subfigure}  {0.35\linewidth}
\includegraphics[width=1\linewidth]{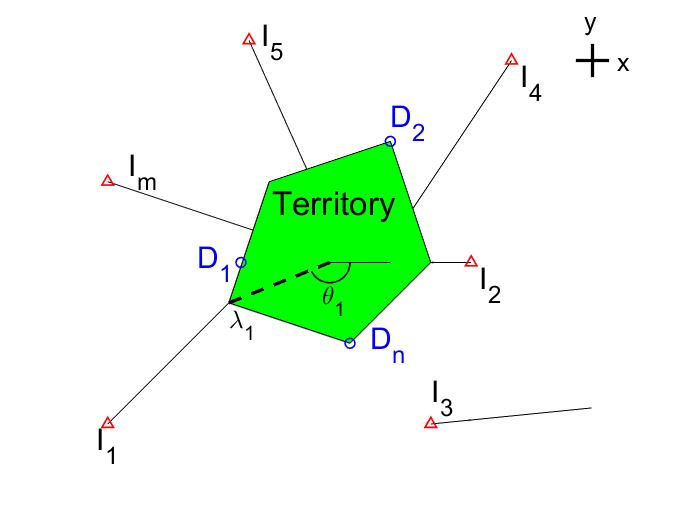}  
\caption{Estimated location of intrusion of intruders for a given time $t$} 
\label{fig:tasks}
\end{subfigure}%
{\LARGE$\Rightarrow $}%
\begin{subfigure}  {0.60\linewidth}
\vspace{2em}
\includegraphics[width=1\linewidth]{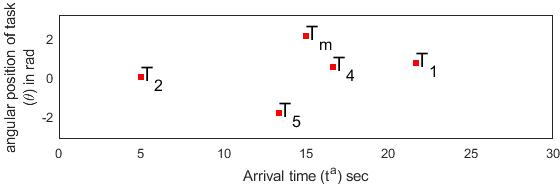}
\caption{Spatio-temporal task for a given time instant $t$ }   
\label{fig:tasksST}
\end{subfigure} 
\caption{A representation of tasks in spatio-temporal space  }
\end{figure*}

In this subsection, we convert the problem of perimeter defense into a Spatio-Temporal Multi-Task Allocation (ST-MTA) problem. Based on the sensor information, the defender will be able to identify the potential intruders. Using the $j^{th}$ intruder position and velocity, the defender can estimate the position $(\lambda_j)$ and arrival time $( t_j^a )$ at which the $j^{th}$ intruder enters the territory. The defender can execute the neutralization task $T_j$, if it can reach the location $(\lambda_j)$ at the time $( t_j^a )$. Hence, the problem is converted into a Spatio-temporal task and is denoted as $T_1 (\lambda_j,t_1^a )$.  The arrival time is computed as  
 
 \begin{equation}
     t_j^a =  \frac{(I_j^s - \lambda_j )}{   V_I^j}
 \end{equation}

 Figure \ref{fig:SenseTask} shows the possible region of intrusion based on the restriction of intruder dynamics. Since the defenders know only the current position and velocity, they will determine a unique point in the perimeter and time of intrusion, as shown in Figure \ref{fig:tasks}. The corresponding spatio-temporal task is shown in figure \ref{fig:tasksST}. The point of intrusion for intruder $j$ is denoted by $\lambda_j$. Now this cartesian point $ \lambda_j (x,y) $ is converted to a polar coordinate $(r_j,\theta_j)$ with centre at the centroid of the territory. The angular position $\theta_j$    uniquely denotes the   $\lambda_j$ on convex territory. For easy understanding we use this angular position $(\theta)$ to denote the location of task in figure  \ref{fig:tasksST} .  

From fig \ref{fig:tasksST}  the task is available at a specific location ($\theta$) at a time $t^a$. This means the task has a strict requirement that is available only at a specific location and at a specific time. If the defender reaches the location $\lambda$ earlier, then it must wait for the task. If the defender reached late at location $\lambda$, the intruder had come early and intruded the territory, and the defense failed. The intruder $I_3$ is not directed towards the territory and hence task $T_3$ is not defined in fig \ref{fig:tasksST}. As intruder changes the heading angle the task changes but smooth heading variations by intruder leads to a smooth variation in task and resolving the task allocation at every time instant can handle the dynamic intruders.

The set of spatio-temporal tasks is denoted as    $ \mathcal{T}=[T_1 (\lambda_1,t_1^a ), \cdots ,T_j (\lambda_j,t_j^a ),\cdots,T_{M_t} (\lambda_{M_t },t_{M_t}^a )] $.  Note, $M_t>N$, defenders will be assigned to multiple tasks at any given time $t$. Let $ {\mu}_i=T_a,T_b,T_c$ be the multiple tasks assigned to the $i^{th}$ defender. Based on the assigned task's arrival time and location, the defender will generate the potential motion commands to follow the trajectory $( {\mu}_i)$. The trajectory $ {\mu}_i$ means that $D_i$ executes tasks $T_a  ,T_b,T_c$ in sequence. 
 
Next, we will discuss the decentralized spatio-temporal multi-task allocation approach to solve  the perimeter defense problem.

\section{ A Decentralized Approach for Multi-UAV Spatio-Temporal Multi-Task Allocation (DMUST-MTA) } \label{sec:DMUST-MTA}
In the previous section, the perimeter defense problem was formulated as a multi-UAV spatio-temporal multi-task allocation problem.  Due to partial observability conditions, all the tasks may not be visible to a defender. A decentralized approach is presented here to solve the MUST-MTA problem.  We are proposing a decentralized solution approach in which each spatio-temporal task cost is computed by combining both the temporal and spatial costs using a composite cost function. The auction algorithm uses this composite cost to compute a local optimal trajectory for each defender. The locally computed trajectories may not be feasible in a global sense. A consensus algorithm is then used to get the globally feasible trajectories.      

\subsection{ Consensus-based Decentralized Approach }
In  section \ref{sec:STMTA for PDP}, the perimeter defense problem was formulated as a multiple UAV based spatio-temporal multi-task allocation (MUST-MTA) problem. Using the sensor information, the defender UAVs can identify the potential intruders who will attempt to enter the convex territory at location $\lambda_j$  and at time $t_a^j$. The tasks  $ T_j$ s of  neutralizing the intruders are   first converted into  spatio-temporal tasks $T_j (\lambda_j,t_j^a )$. For the sake of simplicity, in rest of the paper,   $T_j (\lambda_j,t_j^a )$  is denoted by $T_j$.  
 
In order to protect the convex territory, the defenders will be assigned multiple spatio-temporal tasks in a decentralized fashion, which require the defenders to navigate in a sequence to specific locations  ($\lambda $) at specific times ($t^a $) to neutralize the intruders. The consensus-based decentralized approach for solving MUST-MTA is referred to as DMUST-MTA. The objective of DMUST-MTA is to find the optimal sequence of defenders' motions such that the defenders collectively capture the intruders. The sequence of defenders' motion is a trajectory to be followed by a defender to neutralize the intruders. 
 
 Let us consider that the defender $D_i$ can detect only a few intruders at $ t $ and are denoted by the set $\mathcal{O}_i$. Let the total number of detected tasks in the set $\mathcal{O}_i$ be $M_t^i$. Note that the union of all such sets is equal to the total number of the tasks at any given time $t$, i.e., $\bigcup_{i \in \mathcal{I} }  \mathcal{O}_i   =\mathcal{T} $.   The sequence in which the agent $i$ executes the spatio-temporal tasks is $ {\mu}_i$, where  $ {\mu}_i \in \{ { \mathcal{O}_i} \bigcup  \varnothing \}^{M_t^i }$.  Due to the partial observability condition and overlapping search regions, some intruders are observed by many defenders. Hence, the defenders need to form a consensus among themselves to determine the multiple tasks assigned to them. To form a consensus in this spatio-temporal task allocation problem, the defenders need to estimate the cost associated with each task. The multiple spatio-temporal tasks are then assigned such that the overall cost to the team of defenders is minimum. 
 
 
 Next,   the cost functions for the spatio-temporal tasks consisting of both a spatial and a temporal loss function is described
 
\subsection{ General Cost Function Formulation }
 
Let the available tasks for the  $i$th defender be $\mathcal{O}_i $. The defender $ D_i $ finds a locally optimal sequence ($  {\mu}_i $) to neutralize $\mathcal{O}_i$. The current position of the defender $D_i $ is $ D_i^s$.  At the beginning, the optimal sequence $  {\mu}_i $ is set as  $ \varnothing $  . To find the optimal sequence based on the available task $\mathcal{O}_i$,  it is necessary  to define a composite cost function that combines both  the spatial component and temporal component of the task $T_j $ as given below.

\medskip
\subsubsection*{ \textbf{Spatial loss function}  {$ L_i^s (T_j)$ }} 
The spatial cost for neutralizing the task $T_j  $ depends on the distance to be traveled by the defender $D_i $ from the location $D_i^s ( {\mu}_i,j)$ 
 (Location of the task previous to $T_j $ on trajectory $ {\mu}_i$) to neutralizing point $\lambda_j $, and distance intruder $ I_j  $  needs to travel to reach the perimeter.  the spatial loss function is defined as
 \begin{equation} \label{eq:L_s}
     L_i^s (T_j )= { \| \lambda_j-D_i^s ( {\mu}_i,j)\|}_2+\eta { \| ( \lambda_j   -  I_j^s )     \| }_2
\end{equation}

Where, $ \eta \in (0,1)$ is the scaling factor; the second component reflects the intruder's time to reach the perimeter. For a trivial case, if the defender is already at a location $\lambda_j$, the first term becomes zero, then the second term provides the possibility for the defender to neutralize another task and does not force to wait for task  $T_j$. Consider a defender $D_i $  plans trajectory $ {\mu}_1=T_1,T_2  $ then  the spatial loss is computed as 
\begin{align}  
     L_i^s (T_1 )= { \| \lambda_1-D_i^s  \|}_2+\eta { \| ( \lambda_1   -  I_1^s )     \| }_2    \\
     L_i^s (T_2 )= { \| \lambda_2- \lambda_1 \|}_2+\eta { \| ( \lambda_2   -  I_2^s )     \| }_2
\end{align}

\medskip
\subsubsection*{ \textbf{Temporal loss function}  {$ L_i^t (T_j )$ }   }
Before computing the temporal cost, the defender checks the feasibility of reaching the location $\lambda_j $ within the arrival time of the task ($ t_j^a$). The time spent by the defender $D_i$ for the previous task in a sequence $ {\mu}_i $ be $t_i^a ( {\mu}_i,j) $. The task is said to be feasible if and only if the following    conditions are satisfied  
\begin{align} 
t_j^a  -t_i^a ( {\mu}_i,j ) & >0   \\
 	\frac{ {\| \lambda_j-D_i^s ( {\mu}_i,j) \|}_2 }{ V_D^{max}  } & <(t_j^a  -t_i^a ( {\mu}_i,j ))    
\end{align}

The temporal loss function $ L_{ij}^t ( {\mu}_i )$   defined for agent $i$, to execute task $T_j$ at time $t_j^a$ is
\begin{equation} \label{eq:L_t}
      L_i^t  ( T_j )= 
      \begin{cases}
      \ t_j^a  (t_j^a-t_i^a ( {\mu}_i,j))  &\text{ if task $j$ is feasible}  \\
      \infty  &\text{ if task $j$ is infeasible}  
      \end{cases}
 \end{equation}
Where $ t_i^a ( {\mu}_i,j)=0 $   for the first task on a trajectory; as $T_j $ is the first task in trajectory $ {\mu}_i $ defender $ D_i $ directly starts executing assigned to task  $ T_j $.  

Consider a defender  $D_1 $  plans trajectory $ {\mu}_1  = \{T_1  ,T_2 \}$ then the temporal loss is computed for feasible trajectory 
\begin{align*} 
 L_1^t (T_1 )=t_1^a (t_1^a  -0) \\
 L_1^t (T_2 )=t_2^a (t_2^a  -t_1^a  )
\end{align*}

\medskip
\subsubsection*{ \textbf{ The composite loss function}   $( L_{ij} )$   }

The more general composite loss function can be   defined for a spatio-temporal task as,
\begin{equation}
       L_i ( T_j )  =  f(L_i^s (T_j )  ,L_i^t (T_j )  )    
 \end{equation}
For the PDP protection problem, the function $f$ is selected here is   a multiplication function 
  \begin{equation}  \label{eq:L}
       L_i ( T_j )=  L_i^s (T_j )  .L_i^t (T_j ) 
\end{equation}

It may be noted that the spatial loss function and temporal loss functions are defined for only task $T_j$ and its previous task and not for the entire trajectory.  Also, the loss function does not involve all the tasks carried-out by the defender $ D_i  $. First, we compute the total loss function for the trajectory $\mu_i$ as the summation of all the composite loss functions in the trajectory.
  \begin{equation}  \label{eq:Lsum}
        L_i ( {\mu}_i )   =  \sum_{j \in  {\mu}_i} {L_i  (T_j ) }    
  \end{equation}
  
It should be noted that this total loss function for the trajectory does not consider the sequence in which tasks are done; it just accumulates the loss values for each of the tasks present in the trajectory.  Hence a general cost function is defined for a task $j$ in the trajectory   $   {\mu}_i$. The general cost function $C_{ij}\left(\mu_i\right)$   is based on the effective position of task $j$ in the trajectory $   {\mu}_i$  and is given by,
 \begin{align} \label{eq:cost}
    c_{ij}(  {\mu}_i)&= 
  \begin{cases} 
     \min_ { n \le  \vert  {\mu}_i\vert }  L_i^{ {\mu}_i \oplus_n \{j\} }  - L_i^{ {\mu}_i } & \text{if $ j \not \in   {\mu}_i $ }\\
     \infty   & \text{if $ j \in   {\mu}_i $ }  
  \end{cases}   
 \end{align}

$ \|\cdot\|$ is the cardinality of the set, and $ {\mu}_i \oplus_n j $ denotes that task $j$ is added after  $n^{th}$   element trajectory $   {\mu}_i$. As the task $j$ is added at any location, the new task's cost added to the trajectory is the difference between the total loss function of the new trajectory and the original trajectory. If task $j$ is not in the trajectory, then the cost value is $ \infty  $. 

The multi-task allocation problem is formulated using this cost function, and a decentralized algorithm for task allocation is provided next.

\subsection{Decentralized Spatio-temporal Multi-task Allocation Problem}
The spatio-temporal multiple task allocation problem aims to assign each of the $ M_t  $ tasks to the available $N$ agents such that a task is assigned to only one agent. A defender can execute only the detected tasks. An agent i has detected a few tasks $\mathcal{O}_i $ then can execute tasks present in $\mathcal{O}_i $.        The cost of assigning a task $ j $ to the agent $i$ is $ C_{ij}  $. The tasks are allocated such the cost is minimized.

 The task assignment problem is defined as 
\begin{subequations} 
  \addtocounter{equation}{-1}
\begin{align} \label{eq:MTA}
     \min_{\delta_{ij}}   \quad & \sum_{i=1}^{N }  \sum_{j=1}^{M_t} c_{ij}( {\mu}_i) \delta_{ij}   \\ 
 {\rm such \ that} \qquad  &\sum_{i=1}^{N} \delta_{ij} \le 1\qquad \forall j \in   \mathcal{T}    \label{eq:cost_cond_1}   \\   
 & \delta_{ij} \in \{0,1\}\qquad \forall (i,j) \in  \mathcal{D} \times   \mathcal{T}    \label{eq:cost_cond_2} \\
  & \delta_{ij}  = 0 \qquad if \quad j \notin { \mathcal{O}  }_i     \label{eq:cost_cond_3} 
\end{align}
\end{subequations}

the condition  eq.\eqref{eq:cost_cond_1} enforces that the task can be assigned to only one agent. Eq.\eqref{eq:cost_cond_2} is decision variable $\delta_{ij}$ which states that agent $i$ is assigned to task $j$ or not. Eq.\eqref{eq:cost_cond_3} forces the condition that task can be allocated to defender who has detected the task.

This problem is complex as the cost is a function of the trajectory. This NP-hard problem is solved using two stages. In the first stage, each defender $D_i $  computes its own locally optimal trajectory $\mu_i$. In the second stage from the locally optimal path task is given to the defender who has the lowest cost. Using this, defender fixes their initial trajectory for the next local optimal trajectory generation.

\subsection{Consensus-based DMUST-MTA Approach }

The DMUST-MTA approach is carried out in two stages. First, the local cost-optimal sequence of tasks for every defender is computed.  Each defender computes its optimal cost trajectory. The individually computed may have conflicts with other defenders. More than one defender may approach for a single task. Next, the conflict in the global (team) trajectory is resolved using consensus. The defender with the lowest cost for the task is assigned to the task. The defenders modify the local cost-optimal trajectory based on the allocated tasks over consensus. The defenders will repeat local trajectory computation and global trajectory generation steps until global trajectories are the same as the locally computed trajectories.

 \begin{figure}[!htbp]
 \vspace{-8pt} 
\begin{algorithm} [H]
 \caption{Local trajectory generation by using the composite cost function   (CBBA for agent $i$ at iteration $q$ ) } \label{algo:auction}
 \begin{algorithmic}[1]
\State {\bf{procedure}}   input $ {\bf b}_i(q-1) $, ${\bf \mu}_i(q-1) $, $ {\bf y}_i(q-1) $, $ {\bf z}_i(q-1)$
 \State $ {\bf b}_i(q)  =   {\bf b}_i(q-1) $; 
 \State  $ {\bf \mu}_i(q)  =   {\bf \mu}_i(q-1) $ 
 \State $ {\bf y}_i(q)  =   {\bf y}_i(q-1) $; 
 \State  $ {\bf z}_i(q)  =   {\bf z}_i(q-1) $;
 \State \text{conflict resolved}  $= 0$ \;
 \While {\text{conflict resolved} $= 0 $}
    \% {\textit{Auction Algorithm}   }  \; 
   \State    $c_{ij} = \min_ { n \le  \vert {\bf \mu}_i\vert }  L_i^{{\bf \mu}_i \oplus_n \{j\} }  - L_i^{{\bf \mu}_i }, \ \  \forall j \in {\mathcal{O}^i } \backslash {\bf b}_i $ \;
   \State    $h_{ij} =   {\mathbb I}(c_{ij} < y_{ij}), \qquad \forall j \in \mathcal{O}^i  $ \;
  \State $ J_i = argmin_j    \  c_{ij} . h_{ij} $ \;
  \State $  n_{i,J_i} = argmin_j    \  L_i^{{\bf \mu}_i  \oplus_n \{j\} } $ \;
   \State ${\bf b}_i  = {\bf b}_i  \oplus_{end} {J_i} $ \;
   \State ${\bf \mu}_i  = {\bf \mu}_i  \oplus_{n_{i,J_i}} {J_i} $ \;
   \State ${  y}_{i,J_i}(q) = c_{i,J_i} $\;
  \State ${  z}_{i,J_i} = i $ \;
 \State  \textbf{Call  Consensus Algorithm  }\;
 \EndWhile \; 
\end{algorithmic}
  (Remark: minimization over all $\infty$  value is taken as $\infty$.  All $\infty$ means that the task $j$ is infeasible along path ${\bf \mu}_i $) 
\end{algorithm}  

\vspace{-12pt} 

\begin{algorithm}[H]
 \caption{ Computation of a feasible global trajectory  \\(Consensus by agent $i$ at iteration $q$)  } \label{algo:consensus}
 \begin{algorithmic}[1]
\State {\bf{procedure}}   input $ {\bf b}^k$, ${\bf \mu}^k$, $ {\bf y}^k$, $ {\bf z}^k$  (data received from agent $k$ via synchronized communication ) \;
\If{ $z_{kj}^k = \emptyset $ }
\State $y_{kj}^k = \infty $
\EndIf  
\If{ 
$z_{kj}^k = k $ }
\If{  $z_{ij}^i = k $}
\State Update 
\Else
\If { $y_{kj}  <    y_{ij}   $} 
\State Update
\EndIf
\EndIf
\EndIf
 \If{ $z_{kj}^k = i $ }
\If{  $z_{ij}^i = k $}
\State Reset
 \EndIf
\EndIf
 \If{ ${  z}_{pj}^k = p ;  \ p \neq \{i,k \} $ }
\If{  ${  z}_{ij}^i \neq p $}
\If { ${  y}_{pj} ( = y_{kj}) <    {  y}_{ij}   $} 
\State Reset
\EndIf
\EndIf
\EndIf 
\If {  $  {\bf z}_i  = {\bf z}_k     \qquad  \forall i,k \in {\cal I }  $ } 
\State conflict resolved  = 1
\EndIf
\State Update : $  y_{ij} = y_{kj} , z_{ij} = z_{kj} $ \;
\State Reset : $ y_{ij} = \infty  ,  z_{ij} =  \emptyset $ \;
\end{algorithmic}
\end{algorithm}
\vspace{-6pt} 
\end{figure}

\medskip
\subsubsection*{\textbf{Stage1 : Find the local optimal cost trajectory} }

Let $ \bf{y}_i \in  \mathbb{R}_+^{M_t^i }$ be the vector of winning (smallest) bid value for all the spatio-temporal available at time $t$. For agent $ D_i$ and task, $T_j$,  $ \bf{y} $ is initialized as   $\bf{y}_{ij} (q=0)= \infty $ The trajectory is initialized as null, $  {\mu}_i=  \emptyset  $. 

Firstly, the defender computes the cost for all the observed task from initial position, then chooses the minimum cost and bids for a task with minimum cost as shown in step 8 of auction algorithm. Then agent decides whether the own cost is less than minimum bid for that task. $I(\cdot)$ is an indicator function which is unity for true argument and zero otherwise. If the cost is less than known bid value, the agent bids for the task step 10 in algorithm executes this step. Then agent adds this task to trajectory and initial trajectory is fixed.  After these agents communicate with each other and form a consensus.  The consensus algorithm is explained \ref{algo:consensus}. After having a consensus each agent computes its own trajectory starting from new initial position (the position of last task in trajectory) and computes the cost for the observed task which are not in trajectory

Each agent generates its own local cost-optimal cost trajectory for the tasks. Freezing the local cost-optimal trajectory, we solve the MTA problem.

\medskip
\subsubsection*{\textbf{Stage 2: Global trajectory selection algorithm}}  

Stage 2 solves the MTA problem for the given fixed trajectory $\mu_i$. As trajectories are fixed for each defender are fixed the cost matrix (C) is constant. For the fixed cost matrix, the given MTA problem reduces to finding a minimum cost for each task. The tasks are assigned to a defender who has the lowest bid. The defender will update the vectors for the winning task and winning agent. Some defenders will lose a task from their trajectory. The defender must discard the trajectory from the first lost task as the entire trajectory after the unassigned task is invalid. Multiple defenders may bid for a single task, hence creates conflict between the trajectories. The conflict between multiple defenders for a single task is resolved by finding a minimum bid for that task. The winning bids winning agent and trajectory for the defenders involved in the conflict will be modified.  \ref{algo:consensus} gives a detailed step involved in the global trajectory generation. 

After computing the winning trajectory, winning bids, and winning agent list, each defender, regenerate a locally optimal trajectory. Then the agents share their bid values, and the global trajectory is computed. This procedure is repeated until all local trajectories are unchanged in global computation. In a nutshell, the local trajectories are modified until they become globally valid.

\section{ Performance Evaluation of DMUST-MTA   }\label{sec:results}

The simulations are performed for the perimeter defense problem in a synthetic convex territory region shown in \ref{fig:TP1}. The number of the defenders (N) is set to 3, and the number of the intruder ($N_t$) at time $t$  is limited by 6. The defender can neutralize an intruder within a neutralizing distance (r), and this is set at 5 m. The velocity of all intruders $(v_I)$ is constant and is selected as 3 m/s.  The maximum velocity of all defenders is constrained by $v_D^{max}=4.5 m/s$. 

\subsection{ Case Study }

A total of 6 Spatio-temporal tasks are available at any given time. The results for a single initial condition are presented here; the proposed method's results have been evaluated using Monte-Carlo randomized simulations. The defender's assignment to intruders at a different time is shown in fig \ref{fig:result}.  

The fig \ref{fig:subfig1} shows the initial condition. The defender's sensing region is shown in a shaded region, and the defender can be assigned to the observed intruder.  Fig \ref{fig:subfig1} shows that three defenders can sense  only four intruders. As only four intruders are sensed, three defenders create their trajectory to capture observed intruders. Here we can see that $D_3$ is unassigned.

\begin{figure}[t! ]
     \centering
    \includegraphics[width=0.99\linewidth]{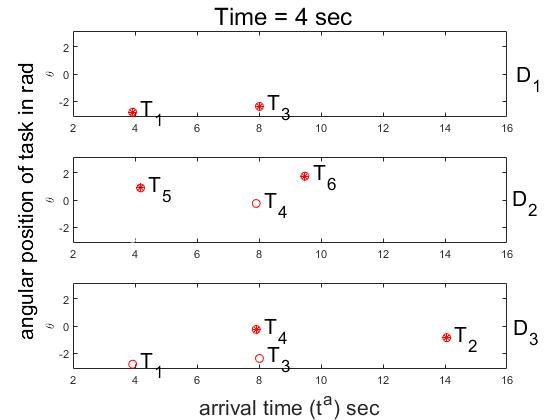}   
    \caption{ Spatio-temporal task allocation in spatio-temporal dimension.   \textmd{The spatial location is denoted by angular position with respect to the center of the territory . $\circ$ shows the observed tasks, and assigned tasks are shown by $\ast$.}}
\label{fig:result_ST_sense}
\end{figure}

\begin{figure*}[htbp!]
\centering
\begin{subfigure}[b]{0.25\linewidth}
\includegraphics[width=1\linewidth]{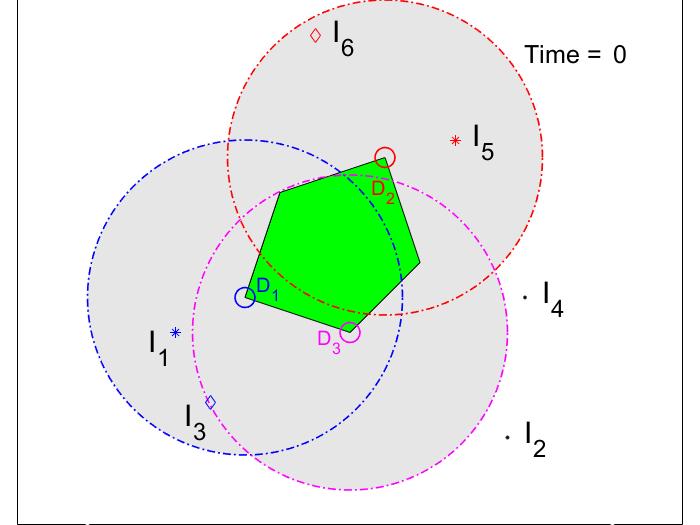} 
\caption{t = 0 sec}  
\label{fig:subfig1}
\end{subfigure}%
\begin{subfigure}[b]{0.25\linewidth}
\includegraphics[width=1\linewidth]{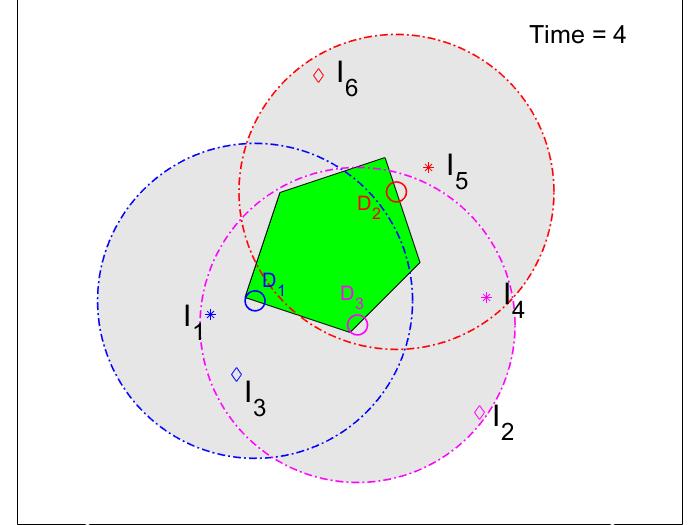}
\caption{t = 4 sec}   
\label{fig:subfig2}
\end{subfigure}%
\begin{subfigure}[b]{0.25\linewidth}
\includegraphics[width=1\linewidth]{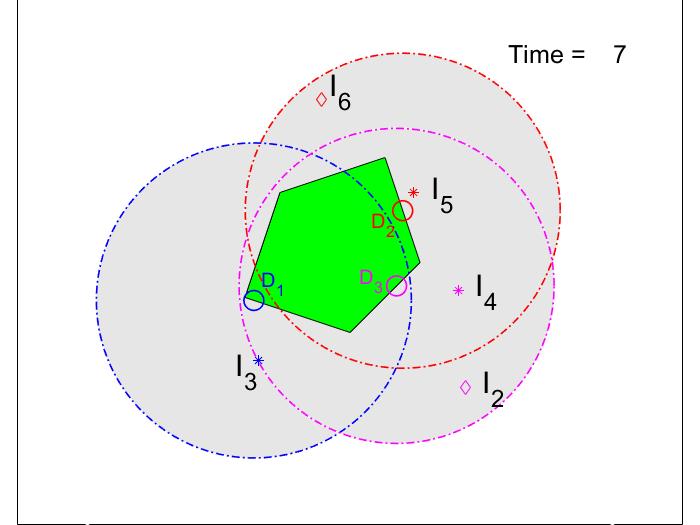} 
\caption{t = 7 sec} 
\label{fig:subfig3}
\end{subfigure}%
\begin{subfigure}[b]{0.25\linewidth}
\includegraphics[width=1\linewidth]{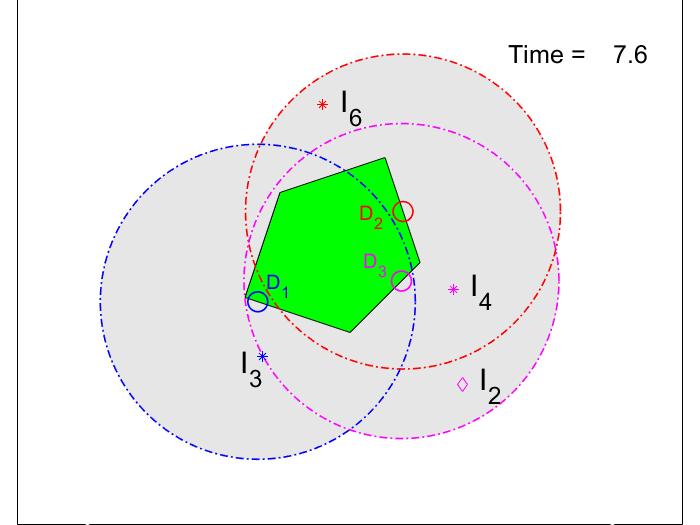}
\caption{t = 7.6 sec}  
\label{fig:subfig4}
\end{subfigure}
\vspace{2pt} 

\begin{subfigure}[b]{0.25\linewidth}
\includegraphics[width=1\linewidth]{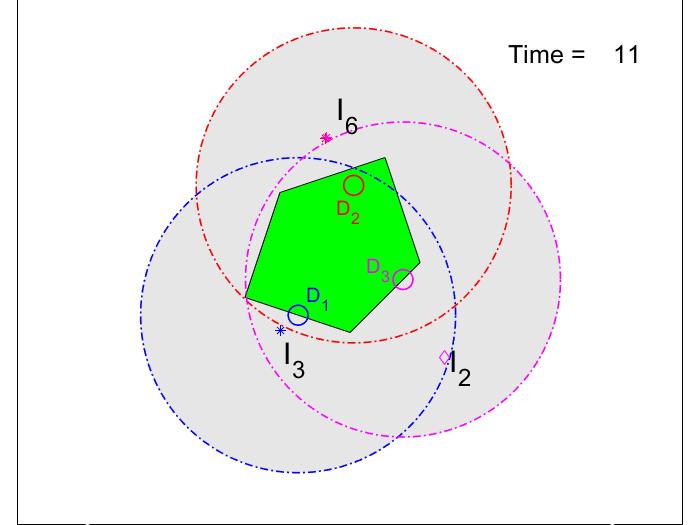} 
\caption{t = 11 sec}  
\label{fig:subfig5}
\end{subfigure}%
\begin{subfigure}[b]{0.25\linewidth}
\includegraphics[width=1\linewidth]{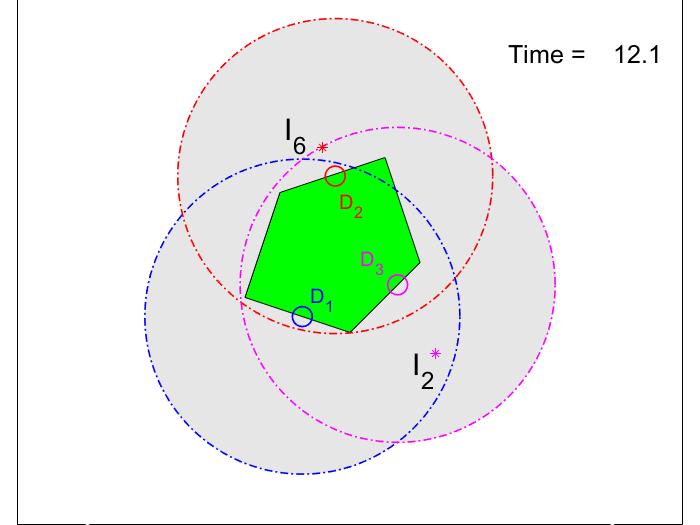}
\caption{t = 12.1 sec}  
\label{fig:subfig6}
\end{subfigure}%
\begin{subfigure}[b]{0.25\linewidth}
\includegraphics[width=1\linewidth]{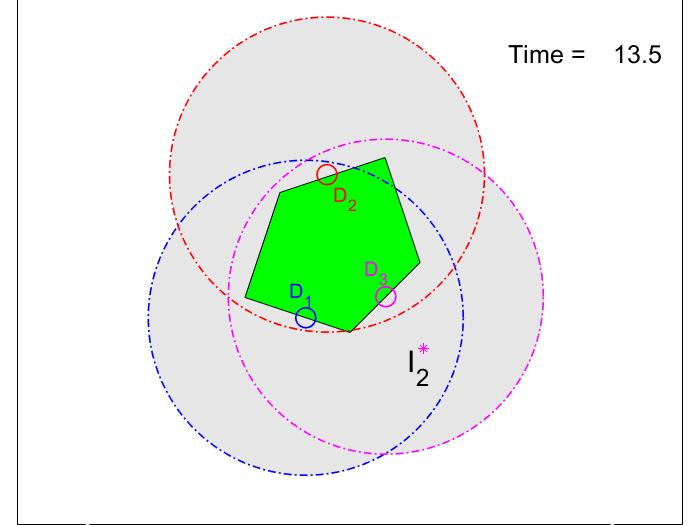} 
\caption{t = 13.5sec}  
\label{fig:subfig7}
\end{subfigure}%
\begin{subfigure}[b]{0.25\linewidth}
\includegraphics[width=1\linewidth]{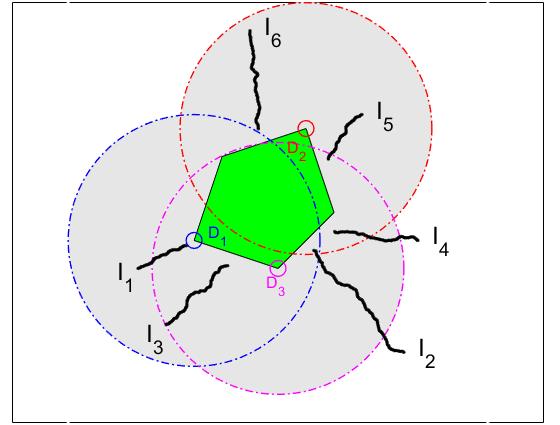} 
\caption{Intruders path}   
\label{fig:subfig_intruder}
\end{subfigure}
\caption{Snapshots of the intruder tasks allocated to the defenders, at different time instants.
\textmd{ Time is given in seconds.  The defenders  $D_1,D_2,D_3$ are represented by colours  blue, red,  magenta respectively.  The intruders assigned to an evader are coloured with the colour of evader. The path selected by evader is shown by symbols in sequence :  *,  $\diamond $, $ \square $, $\lhd$, $\rhd$ .  }  }
\label{fig:result}
\end{figure*}

The fig \ref{fig:subfig2} shows the scenario at 4 sec. defenders can sense all six intruders. The defenders are assigned to intruders n sequence, $D_1$ will neutralize $I_1$ first, and then it will go for $I_3$.  First, the assigned intruder is shown by $\ast$ mark and second by diamond, and third by square. The defender $D_1$ senses $I_1, I_3$ ; $D_2$ senses $I_4,I_5, I_6$; and $D_3$ senses $I_1,I_2,I_3, I_4$. The fig \ref{fig:result_ST_sense} shows the sensed intruder by each defender, and assigned intruders are shown in the dark. The spatio-temporal result shows the time dimension. The defenders will execute assigned tasks as per increasing time order. $D_2$ can sense $I_4, I_5, I_6$; from these three sensed tasks, two tasks $(T_5, T_6)$ are assigned to $D_2$. The sequence can also be observed n time axis of the assigned task where $T_5$ come before $T_6$; hence $D_2$ executes a trajectory  $ \mu_2 = \{T_5, T_6 \}$.

The fig \ref{fig:subfig3}  shows the scenario at 7 sec where $D_1$ has captured the intruder $I_1$, and there are only five intruders. Even though the number of intruders is changed, the same algorithm gives the same previously planned assignment. $D_1$ is assigned to $I_3$, and the remaining assignments are unchanged. The fig \ref{fig:subfig4} shows the scenario at 7.6 sec where, $D_2$ captured $I_5$ and now assigned only to $I_6$. The fig \ref{fig:subfig5} shows scenario at 11 sec, where $D_3$ has captured intruder $I_4$. Now there are three intruders and three defenders. The assignment is the same as planned.  

The \Cref{fig:subfig6,fig:subfig7} shows the capturing of remaining intruders. The intruders are randomly changing the heading angle; each intruder's path is shown in fig \ref{fig:subfig_intruder}.

\subsection{ Performance Evaluation using Monte-Carlo simulations}

Monte-Carlo simulations are performed by varying the speed limits of defenders, sensing radius of the defenders, and inter-intruder time separation. The velocity of the intruder is kept constant, and the maximum velocity of the defender is varied. A total of 15 intruders are considered in a single run, which comes from random positions over a time. At a time instant, at most, six intruders are considered. Success is defined as the capture of all intruders in a run. Even if a single intruder intrudes in the territory, it is considered a failure.

\begin{figure*}[b!]
     \centering
    \begin{subfigure}[b]{0.32\linewidth}
    \includegraphics[width=1\linewidth]{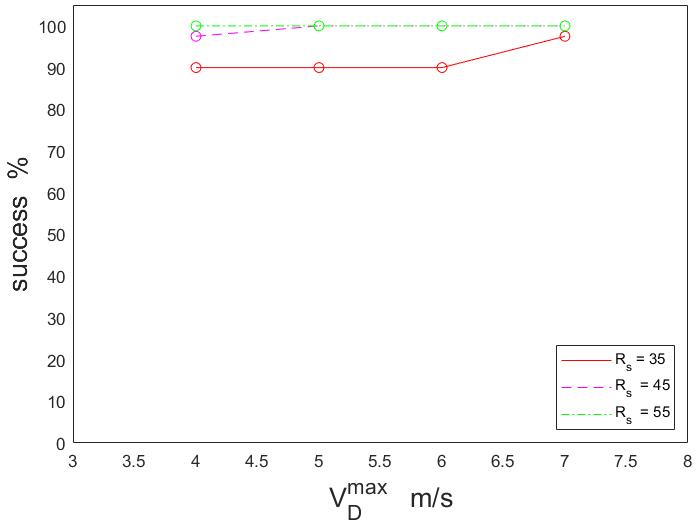} 
    \caption{ $ V_I\ =\ 1m/s $}  
    \label{fig:subfigMC1}
    \end{subfigure} 
    \begin{subfigure}[b]{0.32\linewidth}
    \includegraphics[width=1\linewidth]{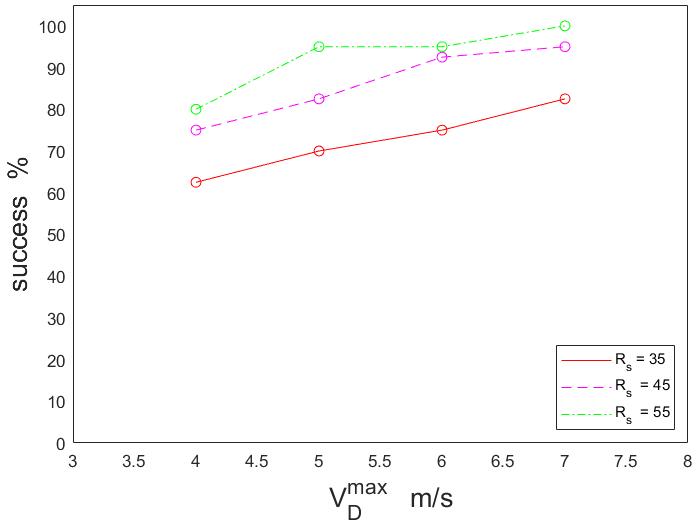} 
    \caption{ $ V_I\ =\ 2m/s $}  
    \label{fig:subfigMC2}
    \end{subfigure} 
    \begin{subfigure}[b]{0.32\linewidth}
    \includegraphics[width=1\linewidth]{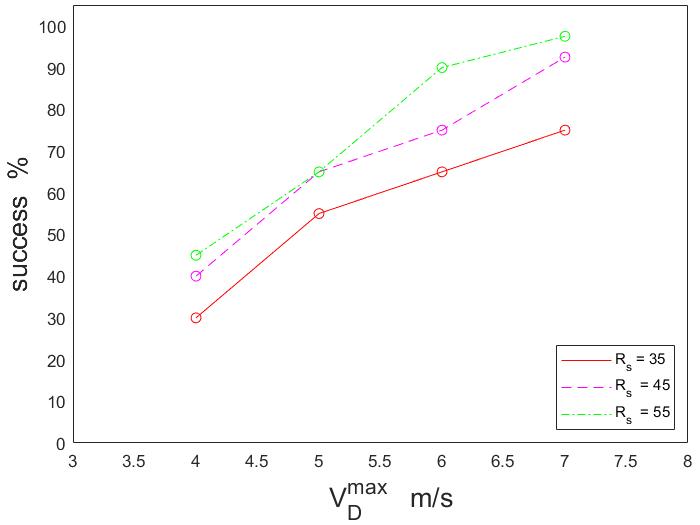} 
    \caption{ $ V_I\ =\ 3m/s $}  
    \label{fig:subfigMC3}
    \end{subfigure}
    \caption{  Variations in defender speed limit      \textmd{$ N = 3 ,  M_t =  6, $ }}
\label{fig:MC_v_D}
\end{figure*} 

The fig \ref{fig:MC_v_D} shows that as the defender's speed limit increases, the success rate to capture intruders increases. The success rate is approaching $100\%$ as randomized simulations may have infeasible tasks due to insufficient temporal separation between consecutive tasks. All three sub-figures \Cref{fig:subfigMC1,fig:subfigMC2,fig:subfigMC3} shows that for a given intruder velocity as the defenders velocity increases the success rate improves. The simulations are conducted with different intruder velocities. The lower the intruder velocity, the success rate is higher. For higher intruder velocity defender needs even higher velocity to capture the multiple intruders. The higher sensing range helps the defender to detect the intruders earlier and plan accordingly. The fig \ref{fig:MC_v_D} shows that the success rate increases with increasing sensing range.

\begin{figure}[hbt!]
     \centering
    \includegraphics[width=0.99\linewidth]{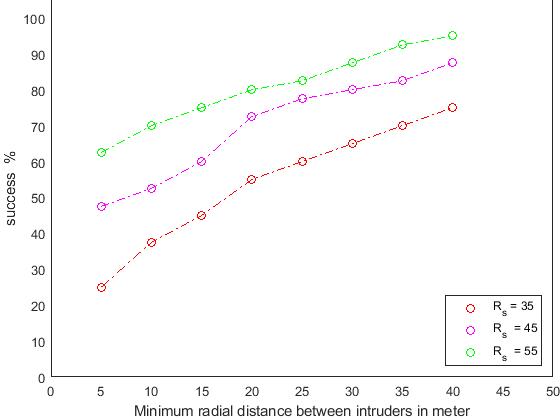}  
    \caption{ Variation in sensing range    \textmd{ $N  =  3,\ \ \ M_t = 6 ,\  \ \ V_I =3m/s \ \ V_D^{max}=6m/s $}}
\label{fig:MC_R_s}
\end{figure}

The next randomized simulations are carried out by varying the inter-intruder time separation. We have used the same set-up for the number of an intruder, except the newly added intruders has minimum time separation in arrival time. This separation time provides feasibility for a smaller number of defenders to capture more intruders. The fig \ref{fig:MC_R_s} shows that as the inter-intruder time separation increases, the success rate increases. If intruders have a temporal separation, then defenders can capture a large number of defenders easily. This helps for strategy making of an intruder; if temporal separation is small, then intrusion rate increases.  Also, the simulations are carried out by varying the sensing radius. As the sensor radius increases, the planning time for the defender increases. The defender can coordinate properly within the defender team, and hence success rate improves. Fig \ref{fig:MC_R_s} show that the success rate increases with increasing sensing radius.

\subsection{ Performance Comparison}

Since the local game region (LGR) approaches present in the literature use a  centralized cooperative perimeter defense game with one defender capturing one intruder \cite{shishika2020cooperative}, we have modified our method to capture one intruder only. Also, the  \cite{shishika2020cooperative} method assumes that the defender has an infinite radius to observe all intruders, and the theoretical solution proposed in those paper are computationally intensive.

\begin{figure}[htb!]
     \centering
    \includegraphics[width=0.99\linewidth]{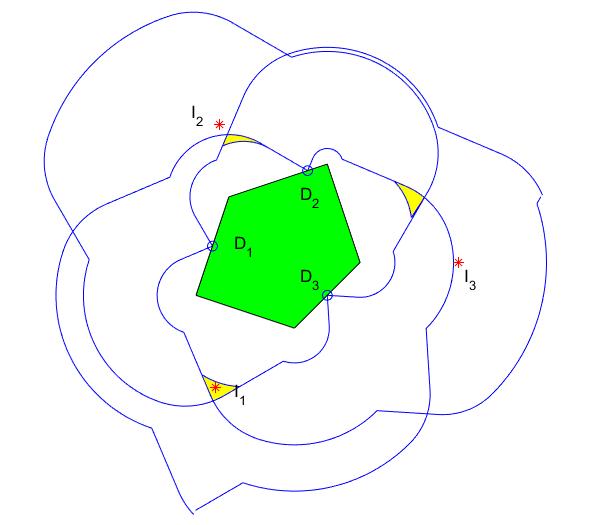} 
    \caption{ Fig: Cooperative perimeter defense game \textmd{The shaded yellow region is a common region for LGR. }}
\label{fig:LGR}
\end{figure}

For computational performance comparison, we have selected the case of three intruders and three defenders. The feasibility region for LGR is shown in the figure \ref{fig:LGR}.  The feasible solution set is
$I_1 \to \{D_1 	\cup D_3 \}$, $I_2 \to \{D_1,D_2 \}$, $I_3 \to \{D_2,D_3 \}$. The solution obtained by the LGR method guarantees at least two capture. The solution obtained is $I_1 \to D_1 	\cup D_3$, $I_2 \to D_2$, $I_3$ is unassigned as of now but may be assigned latter. For the same scenarios, the proposed method allocation is $I_1  \to D_1$, $I_2  \to D_2$, and $I_3 \to D_3$. From the solution, we can also see that both methods guarantee at least two captures. Since both the methods dynamically allocate the task, they may capture all three intruders the time taken by LGR for this allocation 0.3885 sec, whereas the proposed takes 0.008996 sec. The results clearly indicate that the proposed decentralized method is computationally less intensive and can handle partial observability.

\begin{table}[!bht]
\renewcommand{\arraystretch}{1.3}
\begin{tabular}{ |l|r|r|r|r| }
\hline \rule{0pt}{3ex}
& \multicolumn{4}{l|} { Computational time in sec for  number of defenders} \\ [1ex]  \hline
\rule{0pt}{3ex}
 Methods & $N_D$ = 2   & $N_D$ = 3    & $N_D$ = 4   & $N_D$ = 5  \\  [1ex]  \hline    \rule{0pt}{3ex}
LGR  & 0.20308  &  0.388533   &    0.478457  &   0.566431    \\  [1ex]  \hline \rule{0pt}{3ex}
DMUST-MTA  & 0.006958   & 0.008996  & 0.010159   & 0.011127  \\   [1ex] \hline 
\end{tabular}
\caption{Average Computational time required to assign task for fully-observable and one-to-one  scenario  }
\label{table:1}
\end{table}

We have also conducted simulations by varying the number of defenders and computational time to allocate the task. For cooperative LGR, as the number of intruders increases; in the time complexity is polynomial   $O( N N_t^4) $ \cite{shishika2020cooperative}. The CBBA algorithm converges with  time complexity of $O(N)$ \cite{choi2009consensus}.  The proposed method is a modification of the CBBA algorithm and is expected to have a lower computational time. The comparison of the computation time for cooperative LGR method and proposed DMUST-MTA is given in table \ref{table:1}. 
The table shows the proposed method requires less computational time than the LGR method.  

\ifCLASSOPTIONcaptionsoff
  \newpage
\fi

\ifCLASSOPTIONcaptionsoff
  \newpage
\fi

\section{ Conclusions} \label{sec:conclusion}

A new decentralized approach for PDP with distributed sensing, partial observability, and a dynamic number of intruders has been proposed. Here, the PDP is converted into spatio-temporal multi-task allocation (DMUST-MTA) for better scalability and lesser computational complexity. Each observed intruder is estimated to intrude at a spatial location at a specific time. The intruder is viewed as a spatio-temporal task, and the perimeter defense problem is cast as a multi-UAV spatio-temporal multi-task allocation problem. Further, the modified consensus-based bundle allocation handles two-dimensional spatial-temporal multiple-task allocation in a decentralized fashion with lower computational complexity. The Monte-Carlo simulation studies are conducted for various speed, sensor radius, and minimum separation in arrival time, which clearly indicates the decentralized approach is effective at high speed.  The results also show that a large number of tasks can be executed if they have sufficient temporal separation in the arrival time of the intruders. Finally, the comparison with the centralized LGR method with full observability and one-to-one scenario clearly indicates that DMUST-MTA achieves the theory limit with one-order lower computational time. Further work explores a method to find an optimal utilization of defenders to handle any number of intruders.

\ifCLASSOPTIONcaptionsoff
  \newpage
\fi

\bibliographystyle{IEEEtran}
\bibliography{main_bib.bib}

\end{document}